\documentclass[11pt,a4paper]{article}

\usepackage{ascmac}
\usepackage{amsmath}
\usepackage{amssymb}
\usepackage{bm}
\usepackage[footnotesize,bf]{caption}
\usepackage{fullpage}
\usepackage{color}
\usepackage{float}
\usepackage{graphicx}
\usepackage[hidelinks]{hyperref} 
\usepackage[]{natbib}
\usepackage{subfigure}
\usepackage{txfonts}
\usepackage{threeparttable}
\usepackage{wrapfig}
\usepackage[]{multicol}
\usepackage{wrapfig}
\usepackage{authblk}

\usepackage{floatflt}

\title{\large Complete list of the ASTRO-H Science Working Group}
\date{\vspace{-0.5cm}}
\newcommand{\MakeWhitePaperTitle}{
	\begin{center}
		\begin{figure}
			\vspace{1cm}
			\begin{center}
				\includegraphics[width=0.2\hsize]{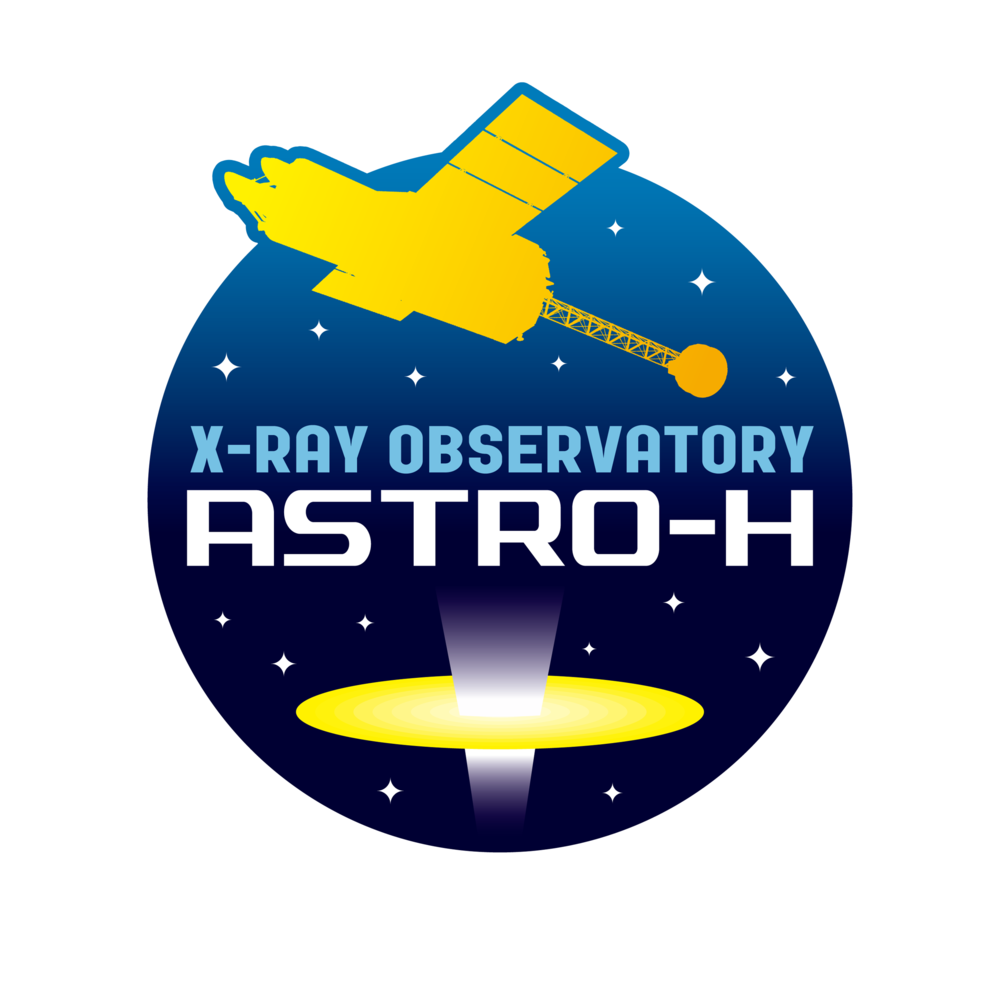}
			\end{center}
		\end{figure}
		\vspace{1cm}
		{\LARGE
		ASTRO-H Space X-ray Observatory\\
		White Paper\\
		}
		\vspace{5mm}
		{\large
		\WhitePaperTitle\\
		}
		\vspace{1cm}
		{
		\WhitePaperAuthors\\
		on behalf of the ASTRO-H Science Working Group
		}
	\end{center}
}

\usepackage{authblk}
\author[a]{Tadayuki~Takahashi}
\author[a]{Kazuhisa~Mitsuda}
\author[b]{Richard~Kelley}
\author[c]{Felix~Aharonian}
\author[d]{Hiroki~Akamatsu}
\author[e]{Fumie~Akimoto}
\author[f]{Steve~Allen}
\author[g]{Naohisa~Anabuki}
\author[b]{Lorella~Angelini}
\author[h]{Keith~Arnaud}
% ...
\affil[a]{Institute of Space and Astronautical Science (ISAS), Japan Aerospace Exploration Agency (JAXA), Kanagawa 252-5210, Japan}
\affil[b]{NASA/Goddard Space Flight Center, MD 20771, USA}
\affil[c]{Astronomy and Astrophysics Section, Dublin Institute for Advanced Studies, Dublin 2, Ireland}
\affil[d]{SRON Netherlands Institute for Space Research, Utrecht, The Netherlands}
\affil[e]{Department of Physics, Nagoya University, Aichi 338-8570, Japan}
\affil[f]{Kavli Institute for Particle Astrophysics and Cosmology, Stanford University, CA 94305, USA}
\affil[g]{Department of Earth and Space Science, Osaka University, Osaka 560-0043, Japan}
\affil[h]{Department of Astronomy, University of Maryland, MD 20742, USA}
% ...

\begin{document}

\newcommand{\WhitePaperTitle}{Chemical Evolution in High-$z$ Universe}
\newcommand{\WhitePaperAuthors}{
	M.~S.~Tashiro~(Saitama~University), D.~Yonetoku~(Kanazawa~University),
	M.~Ohno~(Hiroshima~University), H.~Sameshima~(JAXA),
	H.~Seta~(Saitama~University), H.~Ueno~(Saitama~University),
	T.~Nakagawa~(JAXA), T.~Tamura~(JAXA),
	F.~Paerels~(Columbia~University), N.~Kawai~(Tokyo~Institute~of~Technology),
}
\MakeWhitePaperTitle

\makeatletter
\newcommand{\figcaption}[1]{\def\@captype{figure}\caption{#1}}
\newcommand{\tblcaption}[1]{\def\@captype{table}\caption{#1}}
\makeatother

\begin{abstract}
In this paper, we demonstrate {\it ASTRO-H}'s capability to measure the 
chemical evolution in the high-$z$ $( z \lesssim 3 )$ universe by 
observing X-ray afterglows of gamma-ray bursts (GRBs) and distant 
Blazars.
Utilizing these  sources as background light sources,  the excellent 
energy resolution of {\it ASTRO-H}/SXS allows us to detect emission and 
absorption features from heavy elements in the circumstellar material in 
the host galaxies, from the intergalactic medium (IGM) and in the ejecta 
of GRB explosions. 
In particular, we can constrain the existence of the  warm-hot intergalactic 
material (WHIM), thought to contain most of the baryons at redshift of 
$z \lesssim 3$, with a typical exposure of one day for a follow-up observation 
of a GRB afterglow or 300 ks exposure for several distant Blazars.  
In addition to the chemical evolution study, the combination of the SGD, 
HXI, SXI and SXS will measure, for the first time, the temporal behavior 
of the spectral continuum of GRB afterglows and Blazars over a broad energy 
range and short time scales allowing detailed modeling of jets. 
The ability to obtain these data from GRB afterglows will depend critically 
on the availability of GRB triggers and the capability of {\it ASTRO-H} to 
respond rapidly to targets of opportunity. 
At the present time it seems as if {\it Swift} will still be functioning normally 
during the first two years of {\it ASTRO-H} operations providing the needed 
triggering capability.
\end{abstract}

\maketitle
\clearpage

\tableofcontents
\clearpage


\section{Background and Previous Studies}

In the optical and near IR, deep Hubble and ground based observations 
has revealed the formation and evolution of galaxies up to $z \sim 8$ 
\citep{Bouwens10} allowing constraints  on the star formation rate over 
the last 13~Gyrs of cosmic time. 
While the  star formation rate  is directly connected to the chemical evolution 
of the universe, one must make assumptions about the initial mass 
function, the types of supernova that have exploded and the yields of 
the SN to directly relate star formation to chemical production. 
Direct measurement of the abundance of the elements at high redshifts 
is thus critical but has proven to be difficult, except for a few high 
signal to noise optical observations of the host galaxies of gamma-ray 
bursts. 
Somewhat surprisingly present day observations of the high redshift 
IGM has shown little evidence for chemical evolution.  

Based on models of galaxy formation, the relative population of massive 
stars is thought to increase in the high-$z$ universe and the rate of 
type Ia supernova is thought to be lower than those of type II. These 
assumptions can be checked via measurement of the Fe$_{\rm II}$ / 
Mg$_{\rm II}$ ratio. 
However,  no clear evidence of chemical evolution in 
Fe$_{\rm II}$ / Mg$_{\rm II}$ ratio has been detected and there is a  
fairly large scatter in this ratio seen in the high redshift universe \citep{DeRosa11}. 

X-ray observations, in contrast to optical and IR observations,  can 
determine the abundances  in massive systems such as clusters and groups 
of galaxies and, potentially, in the dominant baryonic reservoir at $z<1$, 
the hot ionized intergalactic medium.
In the last two decades, there has been extensive analysis of the abundances 
in the hot  intra-cluster medium (ICM) at $z<0.5$, showing little if any 
evolution; but the data at higher redshifts are rather poor. 
However, despite of {\it ASTRO-H}'s high sensitivity compared with the 
previous X-ray mission, it will require very long observations to 
measure cluster chemical abundances at high redshifts.

In this white paper we  show the possibility of measuring the chemical 
evolution of the universe beyond $z \sim 1$ utilizing bright background 
light sources such as Blazars and X-ray afterglows of gamma-ray bursts (GRBs).
Since the X-rays are absorbed by primarily K shell electrons, we can 
measure not only cold neutral gas but also the expected warm-hot 
intergalactic medium (WHIM) and the hotter ICM with X-ray absorption
features. Thus the X-ray spectra will measure not only the local 
intra-galactic medium around these sources but also the global chemical 
environments in the high-$z$ universe.

A significant detection of the WHIM has been one of 
critical observational issues in high energy astrophysics, 
since it should provide a solution to the  ``missing baryon problem''.
At $z=0$, the local baryon density calculated from summing up well-
observed baryons  (viz. stars, neutral atomic gas, molecular gas and 
X-ray emitting hot gas) is $\Omega \approx 0.0068$ for 
$H_0=70\ \mathrm{km\,s^{-1}\,Mpc^{-1}}$ \citep{Fukugita98}, 
which is much less than the value expected from Big Bang nucleosynthesis, 
according to the WMAP and Planck observations of the cosmic microwave 
background and what is seen in the high-$z$ universe. 
Thus there must be a large amount of, so far, undetected baryons in the 
local universe consistent with theoretical studies  such as that of \cite{Cen06} 
whose  cosmological hydrodynamic simulations, find that $\sim 50$\% of 
all baryons are in the WHIM at $z=0$. 
This implies that the majority of the missing baryons are hidden in the 
warm-hot material. 
Therefore it is critical to detect it and determine its physical properties. 
The WHIM is expected to be filamentary, shock-heated IGM created during 
the formation  of large-scale structure. 
\cite{Cen06} showed that the mass fraction of WHIM increases from $\sim 
10$\% at $z=3$ to $\sim 50$\% at $z=0$.

For recent decades, several detections of WHIM in the X-ray spectra of blazars
have been reported using {\it Chandra}/LETG or {\it XMM-Newton}/RGS
\citep[e.g.]{Fujimoto04, Williams06, Takei07, Rasmussen07}.
Most of these detections either low statistical significance or cannot confirmed 
with independent observations, except for some exceptions focusing IGM associated
with known signpots of superclusters
\citep{Buote09, Ren14}.
The difficulty in evaluating possible line absorption structures is mainly because
one has to search the entire sightline to rule out statistical fluctuations
\citep{Kaastra06}.

By observing the IGM over a wide range of redshifts, {\it ASTRO-H} might play 
a crucial role in solving the missing baryon problem. 
It is believed that the  WHIM has so far escaped significant detection 
because of the lack of a high throughput, high resolution X-ray spectrometers 
and thus is match to {\it ASTRO-H}'s capabilities, since it is very 
difficult for it to be detected at other wavelengths or by low resolution X-ray spectrometers.
In addition, the {\it ASTRO-H}'s high resolution spectroscopy over a wide energy range
covering iron features, would be powerful to identify the WHIM distinguishing
from hotter intra-group or intra-cluster medium in the line of sight.
Should this search be successful or if sensitive upper limits are obtained, 
it will be a strong observational constraint on the mass fraction and chemical 
evolution of WHIM and thus provide strong impacts on cosmological models 
and star formation histories.

\section{Prospects \& Strategy}

GRB afterglows and Blazars are ideal sources to search for absorption due to  
distant material thanks to their bright structureless continuum, with no intrinsic 
spectral features and their high redshift.
To determine the chemical composition of the absorbing material in the line 
of sight, it is crucial to resolve the fine spectral structures of the edges, 
resonant lines, and emission lines to determine the redshift and their ionization 
state.
It is reasonable to lay a cornerstone in the {\it performance verification phase} 
to evaluate following systematic study of chemical evolution in the high-$z$ 
universe.

The prospects of {\it scientific results} and the strategy for the {\it ASTRO-H} 
early phase observations in this field are:

\paragraph{Investigate the WHIM along the line of sight to well chosen {\it z} $\sim$ 1 -- 3  objects  :}
We propose to employ GRB afterglows and distant Blazars, in order to 
survey the WHIM in the distant universe,  By measuring the redshifts and 
ionization state of the intervening gas, we may reveal the missing 
baryons and their properties. The expected range of the redshift of the spectral features will be 
$1 \lesssim z \lesssim 3$ as discussed in following subsections.

\paragraph{Investigate ISM and eject of Hypernovae with emission 
features in GRB afterglows:}
In addition to the expected absorption features of the IGM, we can also 
search for emission features from  local dense ISM associated with the GRB and from ejecta in 
GRBs afterglow emission.

\subsection{GRB afterglows} \label{sec_intro_GRB}
The progenitor of the ``long'' duration GRB is thought to be due to the core collapse 
of a massive star \citep{Woosley93}. 
Although the redshift distribution of GRBs ranges up to $z \sim 10$ \citep{Jakobsson06}, 
the peaking of the distribution is observed around $z \sim 1$.
Since their radiation energy (of the order of $10^{52}$ erg~s$^{-1}$) is 
$\sim 10$ times larger than those of normal core-collapse supernovae, 
it is widely accepted that the radiation from  GRB emissions is produced in highly relativistic 
jets pointing at the observer.
The  extreme Lorentz factor of $\gamma \sim 10^3$ 
requires a rare type of supernovae as the GRB progenitor.

The tightly beamed high intensity intrinsically featureless emission makes the GRB an 
ideal background light for irradiating intervening matter in the high-$z$ universe.
However, somewhat surprisingly, the  X-ray spectra of high redshift GRBs show evidence for significant 
absorption in excess of the Galactic foreground. the origin of this absorption is not well understood. 
This absorption  is often attributed to the gas in the host galaxy, since the majority 
of long GRBs are associated with the deaths of massive stars and often lie in active 
star forming regions which are associated with dense gas.
If the nature of these star forming regions are similar, the  apparent column density of the X-ray absorbing material should (corrected for the redshift of the source) be roughly constant with 
redshift.
However, detailed analysis \cite{Campana10}, \cite{Behar11}, \cite{Watson12}, 
and \cite{Starling13},shows that many high redshift bursts 
exhibiting high intrinsic absorption beyond that seen in lower redshift objects.
\cite{Behar11} proposed that the X-ray flux from distant objects such 
as GRB afterglows and distant Blazars inevitably suffers absorption 
from the IGM in addition to possible absorption in the host galaxies.
Following \cite{Behar11}, \cite{Starling13} recently showed that 
moderately metal enriched and  warm gas with an estimated ionization 
parameter with $\xi \equiv (L/n_\mathrm{e} r^2) \sim 20$ is a viable expalnation for the observations 
of the high redshift high column density  X-ray absorption, where the $L$, $n_\mathrm{e}$, and $r$ are
the luminosity, electron number density, and distance of the ionized plasma
from the light source.
This WHIM like absorber exhibits a relatively {\it flat} optical depth with 
redshift in the soft X-ray band which seems to compensate for the expected decrease 
of optical depth at the observer rest frame.
These papers clearly showed that X-ray spectoscopic studies of GRB
afterglows are quite promising method to investigate the IGM at high redshift. 
However, in the absence of sensitive high resolution spectroscopy, 
it is impossible to conclusively determine the nature of absorbers 
and their locations especially for the redshift distributions.
The spectral resolution of absorption features is the key to determining 
their chemical abundance, ionization state and physical origin. 
This is undoubtedly one of the frontiers for {\it ASTRO-H} to pioneer.

In addition to the diffuse IGM, the expected dense circumstellar medium 
(CSM) associated with the GRB progenitor may produce  heavy 
elements atomic lines and/or  edges in the afterglow 
spectrum.
So far,  the best evidence for such features as been in the Fe-K band. 
Marginal evidence for iron features has been claimed in several 
X-ray afterglows \citep{Piro99, Yoshida99, Piro00}.
They, however, are still controversial, not only due to the limited statistics 
but also for the rather tight upper limits in other afterglows \citep{Yonetoku00}.

Almost 100~\% of GRBs are followed by the bright X-ray afterglows while
only $\sim 60$~\% show the detectable optical counter parts.
One of the reason for the lack of optical emission is caused by 
the absorption in the host galaxies of GRBs.
Thus these optically dim bursts should show strong X-ray 
absorption signatures making X-ray observations of GRB {\it X-ray}
afterglows a serious contender for detailed {\it ASTRO-H} observations. 

We demonstrate the capability to resolve the expected features in the 
X-ray afterglow in \S~\ref{sec_TS}.1 and \ref{sec_TS}.2.

\subsection{Distant Blazars} \label{sec_intro_Blazar}
Blazars are active galactic nuclei in which the photon emission is dominated 
by a relativistic jet pointing toward observers, and their energy spectra
has less intrinsic spectral features. 
While Blazars are the most luminous AGN, they are not as instantaneously 
bright as X-ray afterglow of GRBs. But their activities last much longer 
than GRBs, and we must enable to perform continuous observations with 
longer exposures. Deep {\it ASTRO-H} observations of Blazars are 
promising prospects to search for intergalactic absorption features in 
their spectra. 

Although the intrinsic Blazar spectrum is thought to be a featureless continuum, 
there were some reports of the detection of absorption spectral features 
in the X-ray of several Blazars.
Using the {\it Einstein} Observatory, \cite{Krolik85} first reported 
line-like absorption features in the soft X-ray spectra of the Blazar PKS~2155$-$304.
Following this discovery, a number of authors have searched for absorption 
features in Blazar X-ray spectra, and there are several claims of  narrow 
absorption lines originating in the IGM, utilizing the {\it Chandra} and {\it XMM-Newton} grating 
spectrometer observations of low redshift ($z<0.3$) Blazars\citep[e.g.][]{Nicastro05}. 
However the existence of these features are still controversial  \citep{Kaastra06}.

Besides the narrow absorption features, \cite{Behar11} utilizing CCD data 
carried out a systematic estimation of the excess absorption in Blazar spectra.
With their direct measurement of optical depth toward a large sample of 
Blazars, they showed that foreground diffuse IGM dominates the absorption 
in high redshift Blazar spectra as well as in GRB afterglows. 
This result, together with the claimed line absorption features, hints at the 
existence of the long sought spectral features from the WHIM.   
Although the redshift and temperature of the absorber  have not been 
accurately determined, due to the limited energy resolution of the CCD 
spectrometers, the detection of such features with the high energy 
resolution and high throughput instruments onboard {\it ASTRO-H}, 
will allow the redshift, temperature and chemical abundance of the 
suggested {\it intrinsic} absorption to be accurately measured.

\section{Targets \& Feasibility}\label{sec_TS}

\subsection{Origin of Excess of Soft X-ray Absorption}
As in \S~\ref{sec_intro_GRB}, an excess of absorption in soft X-ray 
band against to the local Galactic value was reported in several papers 
e.g. \cite{Behar11} and \cite{Starling13}. 
Interestingly, the observed absorption column density continuously 
increases with the redshift of GRBs, which is inconsistent with the simple 
scenario that the absorption is mainly due to the host galaxy of GRBs. 
Therefore, those authors suggest that this excess absorption is due to 
intergalactic material (IGM). 
\cite{Starling13} showed that  the absorption is consistent with it
being due to an ionized absorber 
(the WHIM) with a temperature of 10$^{5-6}$ K and ionization parameter of 
$\xi=20$. 
The estimated excess absorption column density has a mean value of 
10$^{22}$ cm$^{-2}$ per unit redshift. 
Following \cite{Starling13}, we simulated absorption this material using 
 {\tt XSTAR} with parameters of  (gas temperature of 10$^5$ K, 
column density of 10$^{22}$cm$^{-2}$, and the heavy elemental abundance of 0.2), 
and placed at a redshift of 0.1 and with the background continuum due to 
a near and bright GRB afterglow.

\begin{figure}[htbp]
\begin{center}
\includegraphics[width=0.65\hsize]{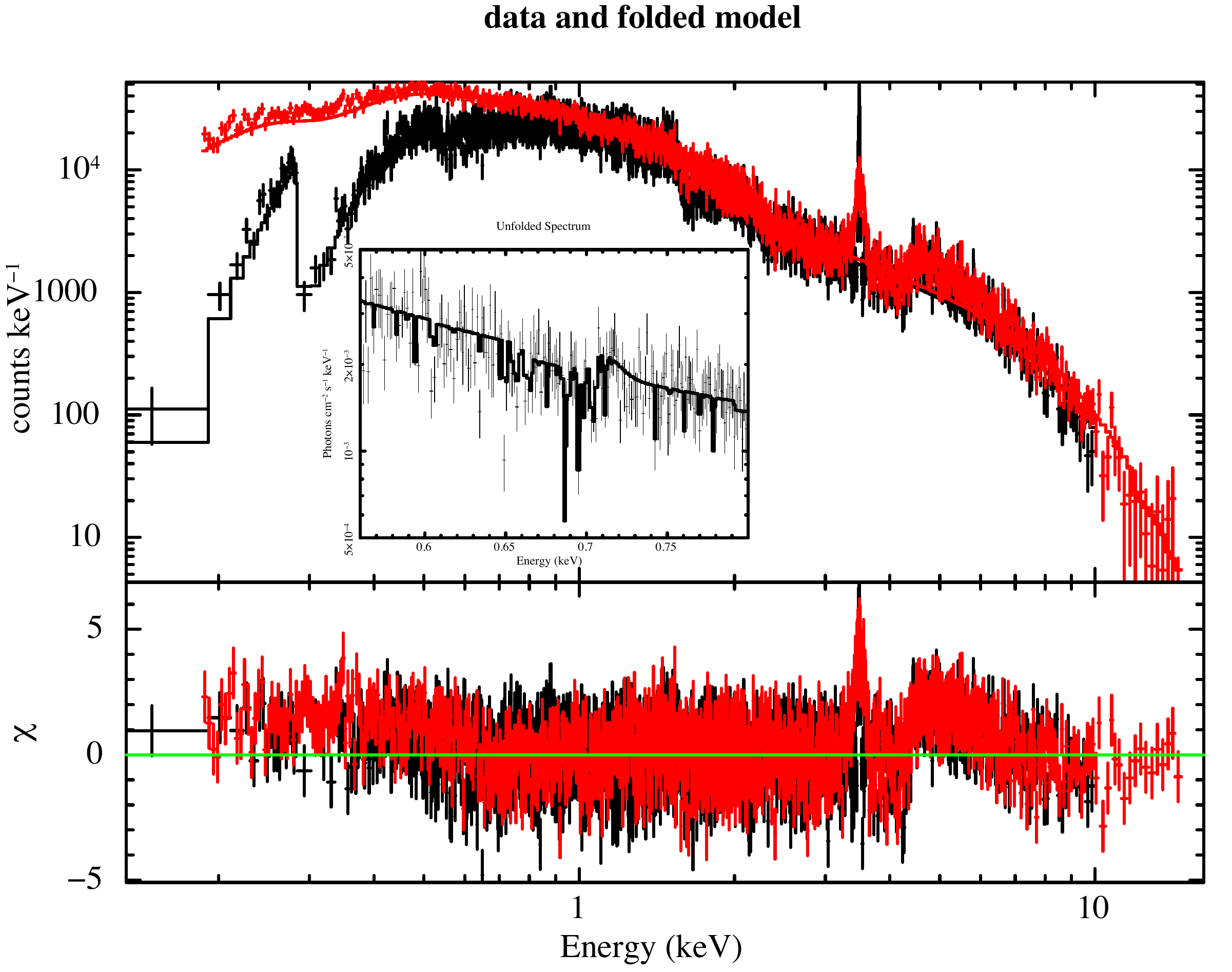}
\caption{
{\it ASTRO-H} simulation of GRB afterglow spectrum with 
the warm absorber model. 
The red points show the SXI and the black the SXS. 
The assumed intrinsic spectral model is taken from \cite{Piro00} as summarized Table~\ref{tab_991216_parameter}, but we assumed narrow line width of 10~eV. 
The absorption in soft X-ray band is replaced to warm material using 
{\tt XSTAR}. 
The redshift of warm absorber is set at $z = 0.1$, the gas 
temperature is $10^5$~K, and the absorption column density, 
$N_\mathrm{H}$ is $10^{22}$~cm$^{-2}$.
A 100~ks exposure is assumed to simulate follow-up observation 
one day after the trigger. 
The enlarged structure around $0.6-0.8$~keV, where 
the most prominent absorption features can be seen are also shown in the 
inset. 
Bottom part 
shows the residuals from single power-law 
model with absorption from cold materials.}
\label{softX-rayabs_inPiro2000}
\end{center}
\end{figure}

A 100ks exposure of a GRB  in which we used as inputs the 
results of \ref{softX-rayabs_inPiro2000}  for GRB~991216 
(as summarized in Table~\ref{tab_991216_parameter}) clearly exhibits 
detections of absorption lines, associated with Fe and oxygen ions 
around 0.686 and 0.695~keV in observer frame with $4-5~\sigma$
significance level. In order to show the sensitivity of 
the SXI and the SXS to the ionization state 
of the intervening gas we also show the residuals from an alternative model 
--- single power-law with absorption by a cold-neutral material --- 
in the bottom panel of the figure.
In addition to these absorption features, we also simulated emission 
features reported in \cite{Piro00} but assumed relatively narrow line case ($\sigma = 10$~eV). 
Detailed discussion on the emission features are described in section~\ref{sec_lines}. 
Such emission line features are clearly detected even for flux level of 
$10^{-12}$ erg~cm$^{-2}$~s$^{-1}$.
Figure~\ref{softX-rayabs_inPiro2000} right shows the 100 ks observation 
of the same afterglow. .
The possibility of {\it ASTRO-H} observing such a  bright GRB afterglow is discussed in detail in \S~3.4.

\begin{table}
    \begin{center}
      \caption{{\it ASTRO-H} simulation of GRB 991216 based on 
\cite{Piro00} results}\label{tab_991216_parameter}
      \begin{tabular}{l|l}
        \hline
        \hline
                     & assumed parameters \cite{Piro00} \\
\hline
        $N_{\rm H}^a$ & 0.35 ($\pm$ 0.15) $\times$ 10$^{22}$ cm$^2$\\
        photon index & 2.2 ($\pm$ 0.2) \\ 
        Edge Energy & 4.4 ($\pm$ 0.5) keV \\ 
        $I_{edge}^b$ & 3.8 ($\pm$ 2.0) $\times$ 10$^{-5}$ ph cm$^{-2}$\
s$^{-1}$ \\ 
        Line Energy & 3.49 ($\pm$ 0.06)keV \\ 
        $I_{line}^b$ & 3.2 ($\pm$ 0.8) \\ 
        Flux$_{2-10 \rm keV}^c$ 
       & 2.3  $\times$ 10$^{-12}$ erg~cm$^{-2}$~s$^{-1}$ \\
        \hline
        \end{tabular}\\
\end{center}
  \end{table}

\subsection{Constrain Progenitor Model and Environment of GRBs by 
Fe-K Line and Recombination Edge}\label{sec_lines}

Spectral emission features in GRB afterglows are a powerful tool to 
investigate the circumstellar environment of GRBs. As described in 
\S~\ref{sec_intro_GRB}, GRB progenitors could be rare type of 
supernovae. \cite{Piro00} reported the 
detection of an Fe-K line feature and a radiative recombination 
edge in the afterglow spectrum of GRB~991216 at 4$\sigma$ confidence 
level. They showed that, utilizing the central energy of Fe-K line 
and radiative recombination edge, we can determine the redshift of 
the GRB as well as provide an estimation of the ejecta mass 
from the supernova.

In order to examine the feasibility, of detecting a similar set of spectral 
features, we simulated the \cite{Piro00} et al data, 
assuming a  power-law with Fe-K line and recombination edge and parameters. 
\cite{Yonetoku00} estimated an upper limit of Fe-K line equivalent width (EW) 
using {\it ASCA} data of GRB~990123 and set it to be 100~eV.  
Accordingly, here we assumed the EW of 50~eV.

\begin{figure}[htbp]
\begin{center}
      \includegraphics[width=10cm]{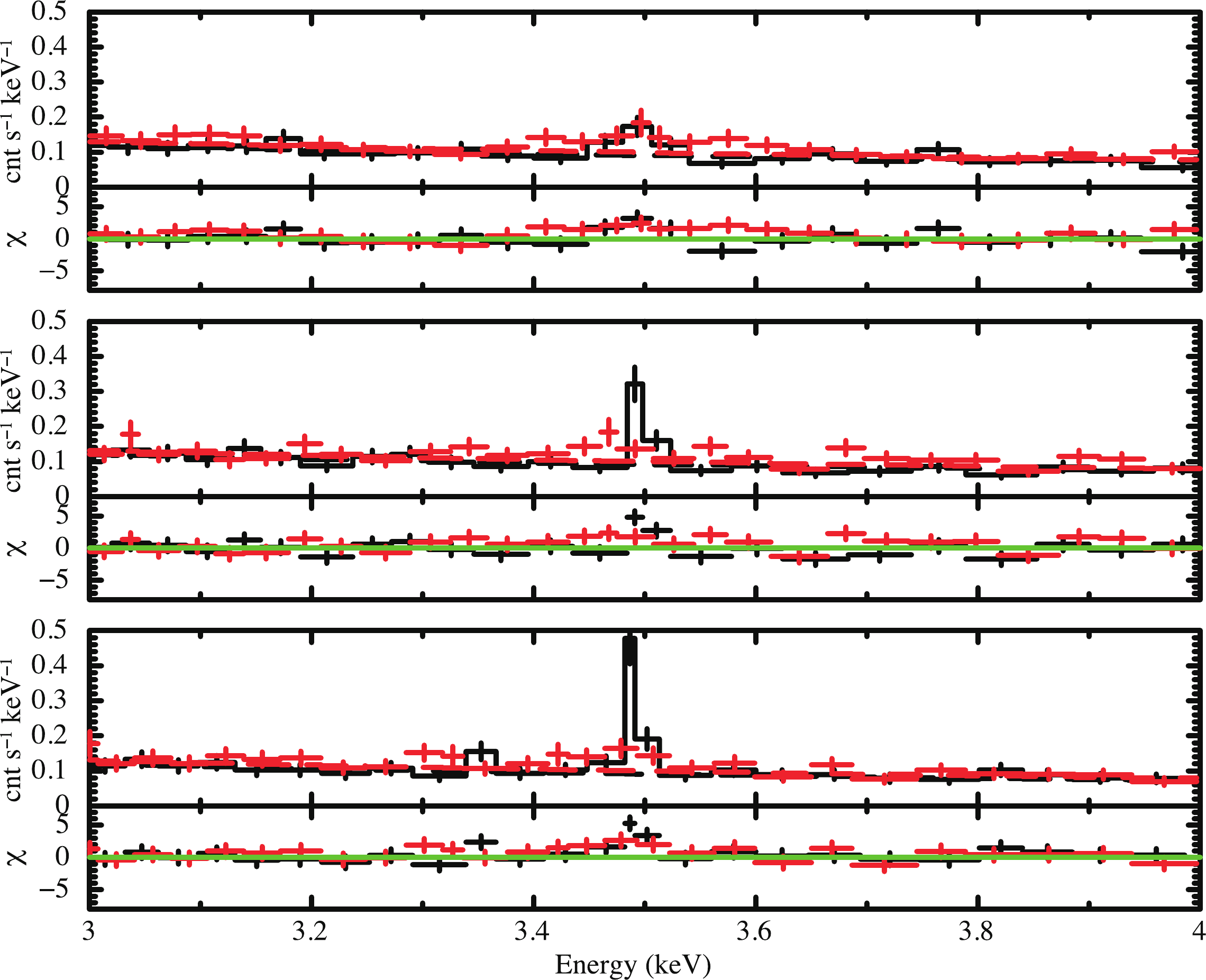}
      \caption{{\it ASTRO-H} simulation of weak Fe-K line features for 
various intrinsic line widths. Each panel contains simulated 10 ks spectrum 
with models (power-law+Fe-K line) and its residuals from single power-
law model. We assumed weak Fe-K lines with equivalent width of EW=50eV 
but assumed bright afterglow spectrum with 
$F_{2-10~\mathrm{keV}} = 10^{-11}$~erg~s$^{-1}$~cm$^{-2}$. 
We used various intrinsic line width 
$\sigma=30$~eV (top), 10~eV (middle), and 5~eV (bottom). }
\label{991216_SXSSXI}
\end{center}
\end{figure}

Measurement of the line width provides crucial information for understanding 
the physics of the ejecta of the GRB progenitor. 
\cite{Piro00} claimed a line width $\sigma \sim 200$~eV, but the observational 
evidence for the width of Fe-K line is not strong due to the energy resolution 
of the CCD data.  
The high resolving power of Fe-K line of SXS is much more sensitive 
than previous instruments for  detecting  weak narrow Fe-K lines. 
In Figure~\ref{991216_SXSSXI}, we show the results of 
simulating Fe-K line emission with the SXS based on the spectra of GRB~991216. 
A weak Fe-K line is detectable by SXS at  5.3$\sigma$ for 5~eV width and 4.8$\sigma$ 
for 10~eV width for a nominal GRB spectrum. 
Such a feature is not detectable with CCD spectral resolution.  

\subsection{Soft X-ray Lines from GRB Afterglow}
In addition to the iron features, emission lines from other light
elements are important in probing the environment of the progenitor 
of GRBs. There have been a few claimed detections of emission lines due
to Mg, Si, S, Ar, and Ca in some GRB spectra \citep{Reeves02, Watson03}. 
However these results have not been confirmed in recent {\it Swift} data 
\citep{Hurkett08}. 
Accurate and significant detection and measurement  of the soft X-ray 
lines with {\it ASTRO-H} SXS/SXI, is needed to confirm the existence 
of such features and their use as indicators of chemical evolution 
in the high-$z$ universe. 
 
We examine the feasibility of detecting the soft X-ray lines 
with {\it ASTRO-H}. In these simulations, we calculated 
the SXS/SXI spectra assuming the reported spectral parameters 
but doubled the X-ray flux.
Since these ISM origin features might be illuminated for 
a short time, we assumed the exposure time of 10 ks in this simulation. 
As we can see in the figure, the soft X-ray line features are clearly 
visible by {\it ASTRO-H} observation with $\sim 4 \sigma$ significance. 
In this spectral simulation, we assume a line width of 5~eV, since no 
significant line broadening was reported in the previous observation. 

\begin{figure}[htbp]
 \begin{center}
\includegraphics[width=10cm]{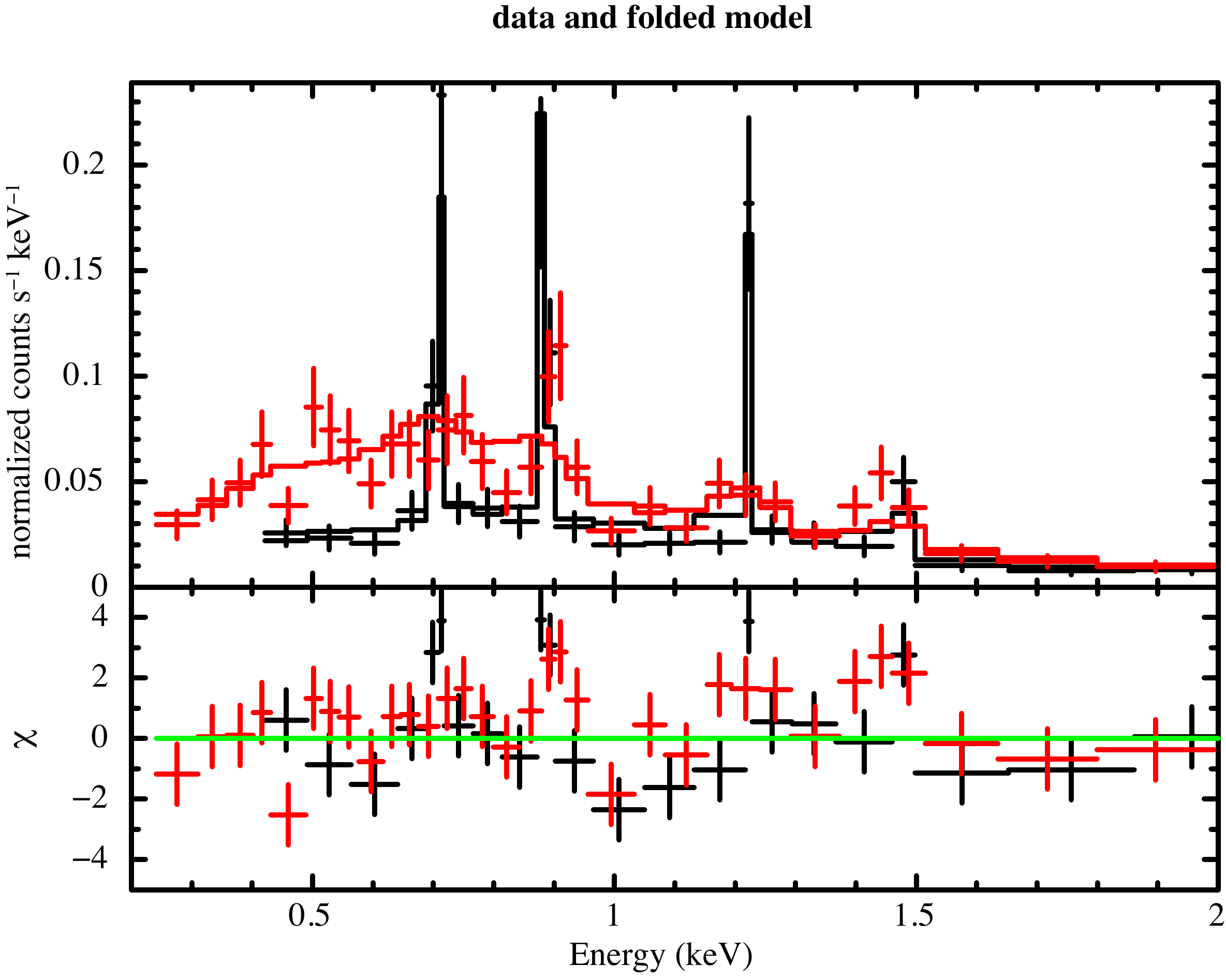}
\caption{
Simulated spectrum of afterglow spectrum of GRB~011211 by {\it ASTRO-H} 
SXS (black) and SXI (red), assuming 10 ks of exposure time. 
Fitted spectral models are also plotted with 
the solid line. The assumed spectral model was taken from \cite{Reeves03} 
but 2 times brighter than previous paper. Bottom panel 
shows the residuals from single power-law model.}
\label{011211_5times}
 \end{center}
\end{figure}

\subsection{Expected Number of Events of GRB Afterglow by {\it Astro-
H}}\label{sec:GRBnumber}
As we showed in previous sections, {\it ASTRO-H} has a capability of 
detecting any spectral features such as Fe-K lines, recombination 
edge, other lines from light metals, and of course absorption by warm 
intervening materials. 
In this section, we estimate the expected number of GRBs 
in which {\it ASTRO-H} can detect these spectral features.
For this estimation, we calculate the luminosity function of X-ray 
afterglow based on {\it Swift}/XRT data base which is publicly available 
in the web site\footnote{\tt http://www.swift.ac.uk/xrt\_curves}. 

\begin{figure}[htbp]
   \begin{center}
     \includegraphics[width=0.55\hsize]{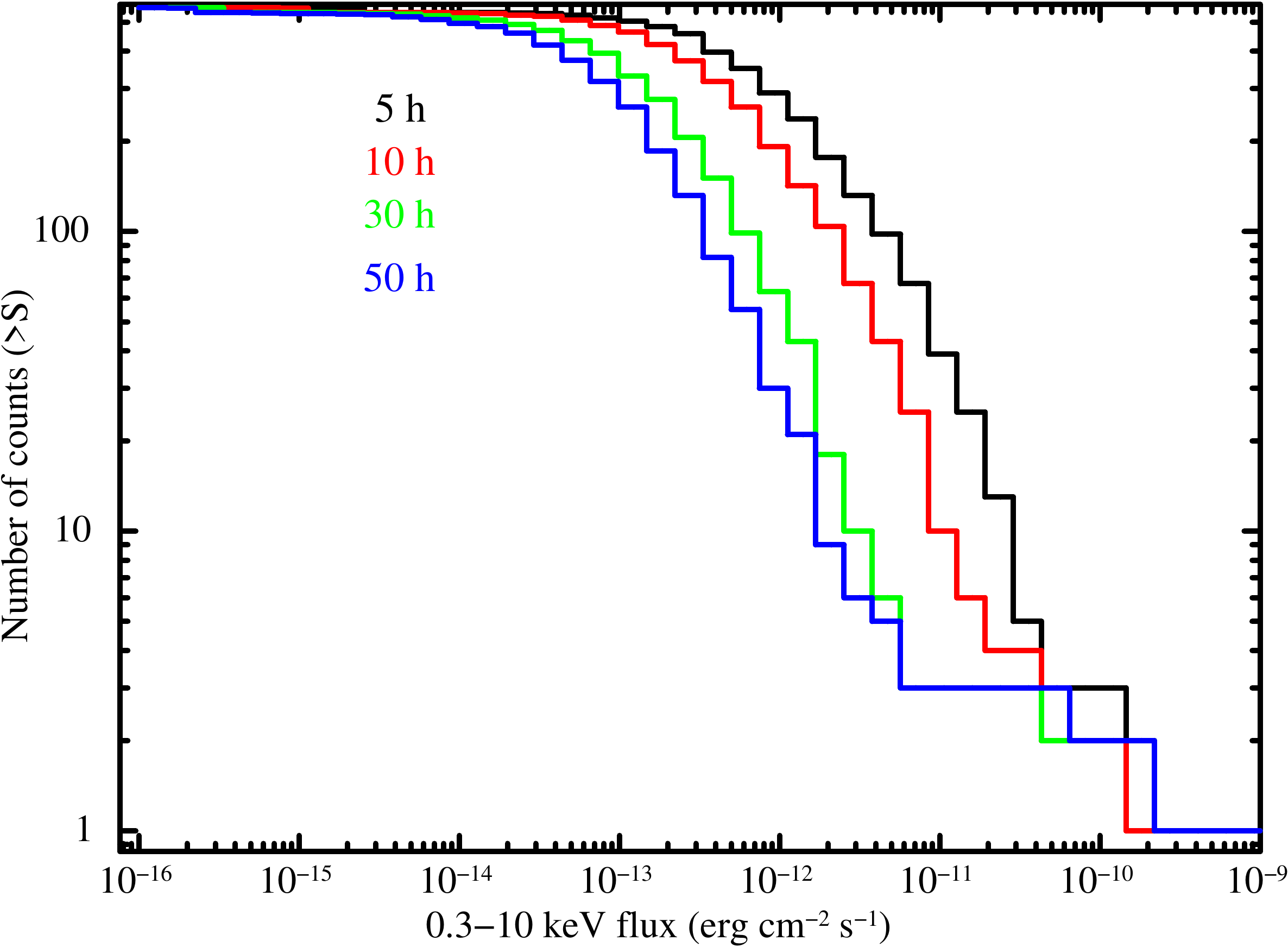}
      \caption{Estimated luminosity functions of GRB afterglow based on 
6-years {\it Swift}/XRT data base. Different colors show the function 
corresponds to the different observation start time after GRB trigger 
(black:5~hour, red:10~hours, green:30~hours, and blue:50~hours, 
respectively). Horizontal and vertical axis is 0.3--10~keV flux in 
erg~cm$^{-2}$~s$^{-1}$ and cumulative number of events, respectively. 
}
\label{LFGRBAG}
\end{center}
   \end{figure}

First, we fit every XRT light curve in the 6-year data base with a 
simple power-law function of time from the GRB trigger time.
We then summarized each afterglow flux of time in ``the cumulative 
luminosity function'' as we showed in Figure~\ref{LFGRBAG}.
Finally, we compared the flux level of simulated spectra in the 
\S~\ref{sec_TS}.1 and \ref{sec_TS}.2 with the estimated luminosity 
function (Figure~\ref{LFGRBAG}).
According to the simulations as previously shown, in the case of 
Fe-K line emission and recombination edge, the flux level of 
$2.3 \times 10^{-12}$ erg~cm$^{-2}$~s$^{-1}$ are required
to constrain the iron spectral features (in case of GRB~991216, 
37~hours after the GRB trigger.
The expected event rate is about 3.3events per
year. On the other hand, in the case of our simulation for detection 
capability of weak Fe-K line, we assumed very bright afterglow with 
the flux level of $\sim$ 10$^{-11}$ erg~cm$^{-2}$~s$^{-1}$. 
Such flux level at late times is rare with $\sim$1 event per year 
10 hours after the trigger and 0.3 event per year for 30 hours after the 
trigger.

In the case to observe the soft X-ray lines, the simulated flux is 
roughly about $10^{-12}$~erg~cm$^{-2}$~s$^{-1}$ level at 11~hours after 
the GRB trigger the expected event rate is about 16 events per year. 
In conclusion, we reasonably expect {\it ASTRO-H} to able to observe 
soft X-ray emission lines from GRB afterglows with an event rate of 
a few per year, even if it takes one day to slew the spacecraft.
But we need to slew within 10~hours after GRB trigger,
to detect Fe-K line and edge features with event rate of at least one 
event per year. 
A follow-up within 10~hours is possible indeed, since 
 {\it ASTRO-H} has a slew speed of over $180^\circ / 100$~minutes.
 However, a dedicated operation team for the quick follow-up
 operation is required as the {\it Suzaku} GRB follow-up team 
 which succeeded carried out four times of quick follow-up observations 
 within a few hours. At this moment, {\it ASTRO-H} team has not decided 
 to organize the quick follow-up observation team. 

Table \ref{GRBnumTable} summarizes feasibility of GRB 
observation by {\it ASTRO-H}.

\begin{table}[htbp]
\caption{Summary of expected number of GRB events for various flux level 
and observation start time}
\label{GRBnumTable}
\begin{center}
\begin{tabular}{l|l}
\hline
\hline
Flux level & event rate per year \\
           & 50 hours / 30 hours ($\sim$ 1 day) / 10 hours ($\sim$ half 
day)\\
\hline
10$^{-11}$ erg~cm$^{-2}$~s$^{-1}$ & 0.5 / 0.5 / 1 \\
10$^{-12}$ erg~cm$^{-2}$~s$^{-1}$ & 3 / 10 / 33 \\
\hline
\end{tabular}
\end{center}
\end{table}

\subsection{Distant Blazar spectral simulation of absorption features by IGM}

Based on their redshift and flux, we have selected four Blazars among 
the object exhibiting significant excess absorption listed in 
Table~1 of \cite{Behar11}. Each of four has a 2 -- 10 keV flux brighter 
than $1 \times 10^{-11}$ erg~cm$^{-2}$~s$^{-1}$ and 
redshift higher than 2 as summarized in Table~\ref{tab:Blazars}. 
In this section, we show the simulated spectrum and the results of evolution, 
utilizing the same {\tt XSTAR} based WHIM model as employed for 
GRB afterglows in \S~\ref{sec_TS}.1 for the representative Blazar RBS~315, 
at a redshift of $z = 2.69$.

\begin{figure}[htb]
\centerline{
      \includegraphics[height=5cm]{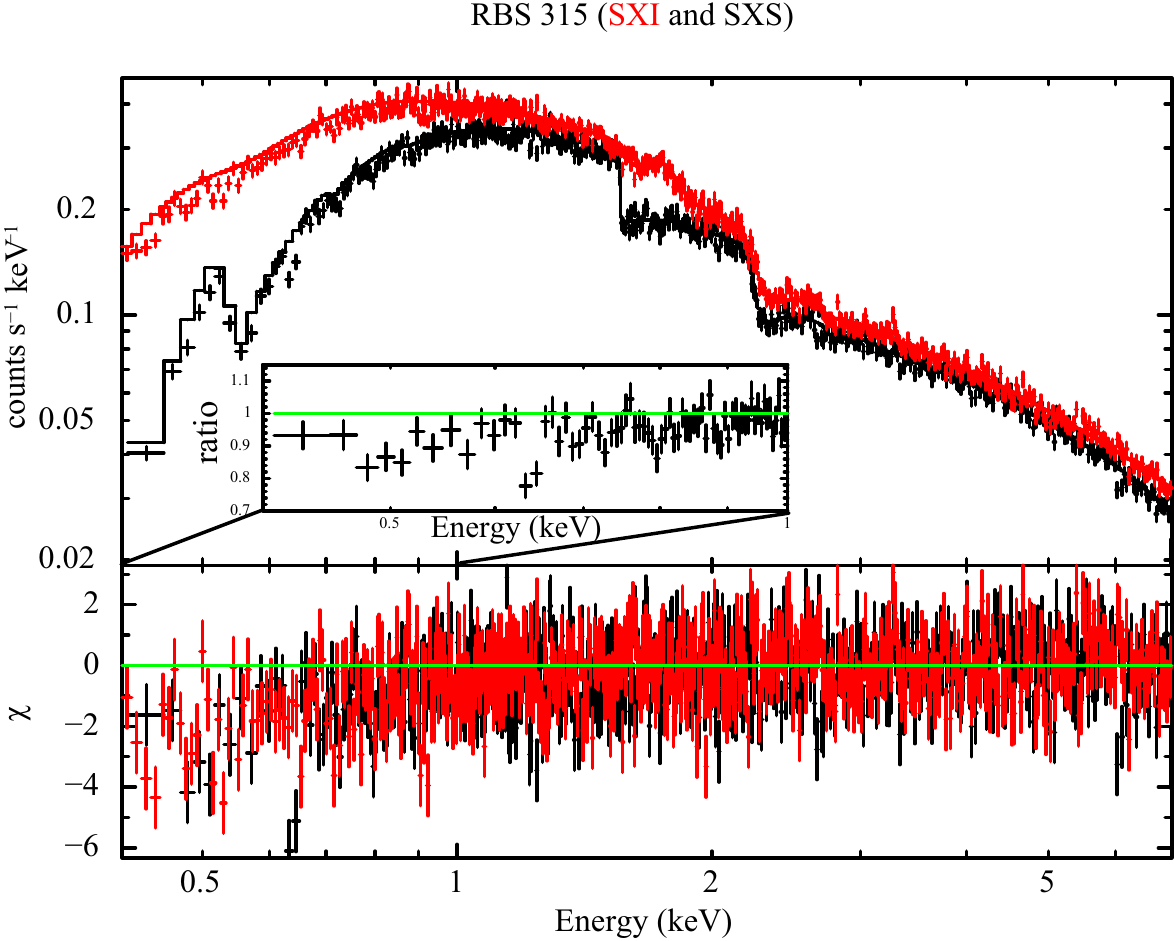}
      \includegraphics[height=5cm]{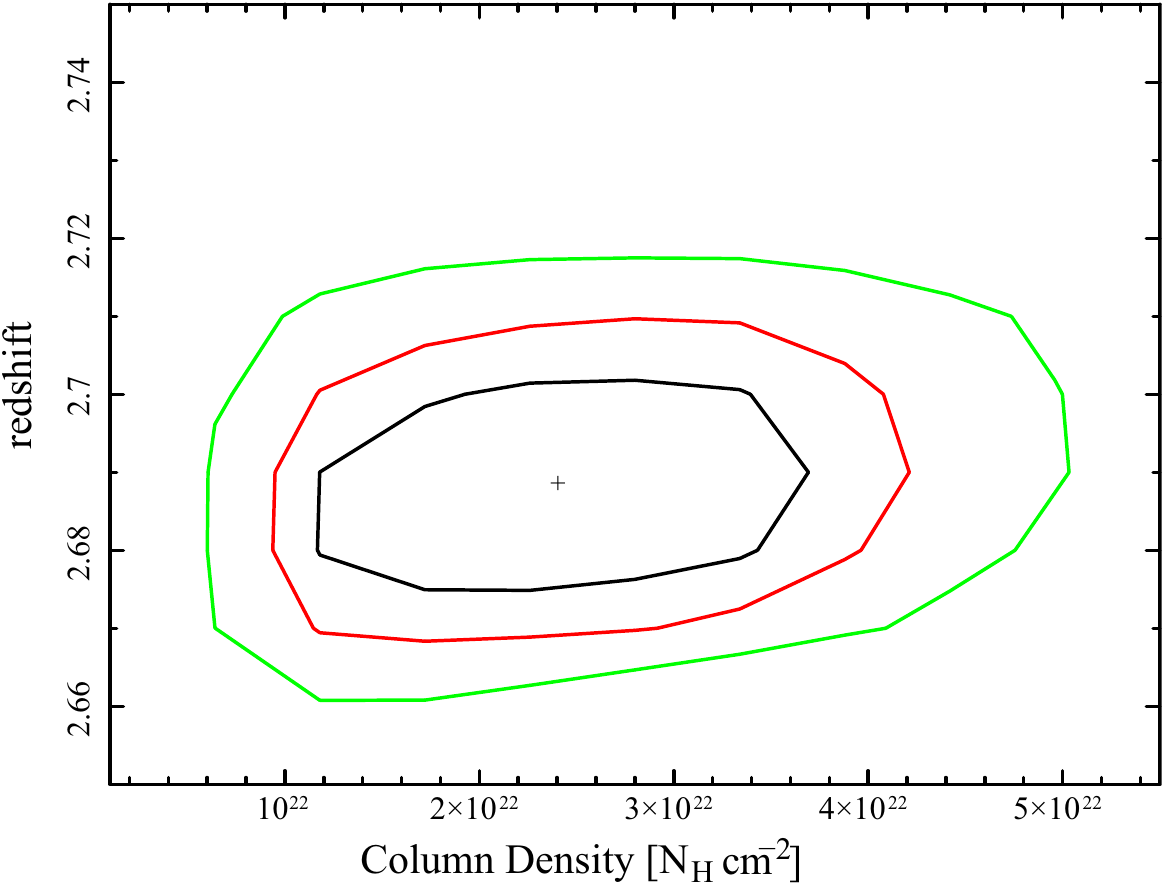}
} 
     \figcaption{Simulated spectrum of the Blazar RBS~315 by {\it ASTRO-H}/SXS (black)  
and SXI (red) according to the result by \cite{Behar11} and a confidence contours 
to evaluate redshift of the {\it intrinsic} absorber.
WHIM (temperature of $10^6$~K, 0.2 solar abundance,column density 
$N_{\rm H} = 2.9 \times 10^{22}$ cm$^{-2}$) absorber at the source rest frame ($z=2.69$) 
is assumed.
The simulated spectrum in the left is shown in crosses, and the model from which the 
WHIM component is removed is indicated with histogram. 
The WHIM originated spectral features are clearly detected as shown in the inset 
data to model ratios in the energy range from 0.4 to 1 keV. Residuals are in the lower panel.
In the right, we show the confidence contours to demonstrate the expected accuracy with a
300~ks exposure. Each contour denotes 68\%, 90\%, or 99\% level confidence.}
\label{fig_RBS315_redshift}
\end{figure}

\begin{figure}[htb]
\centerline{
      \includegraphics[width=7cm]{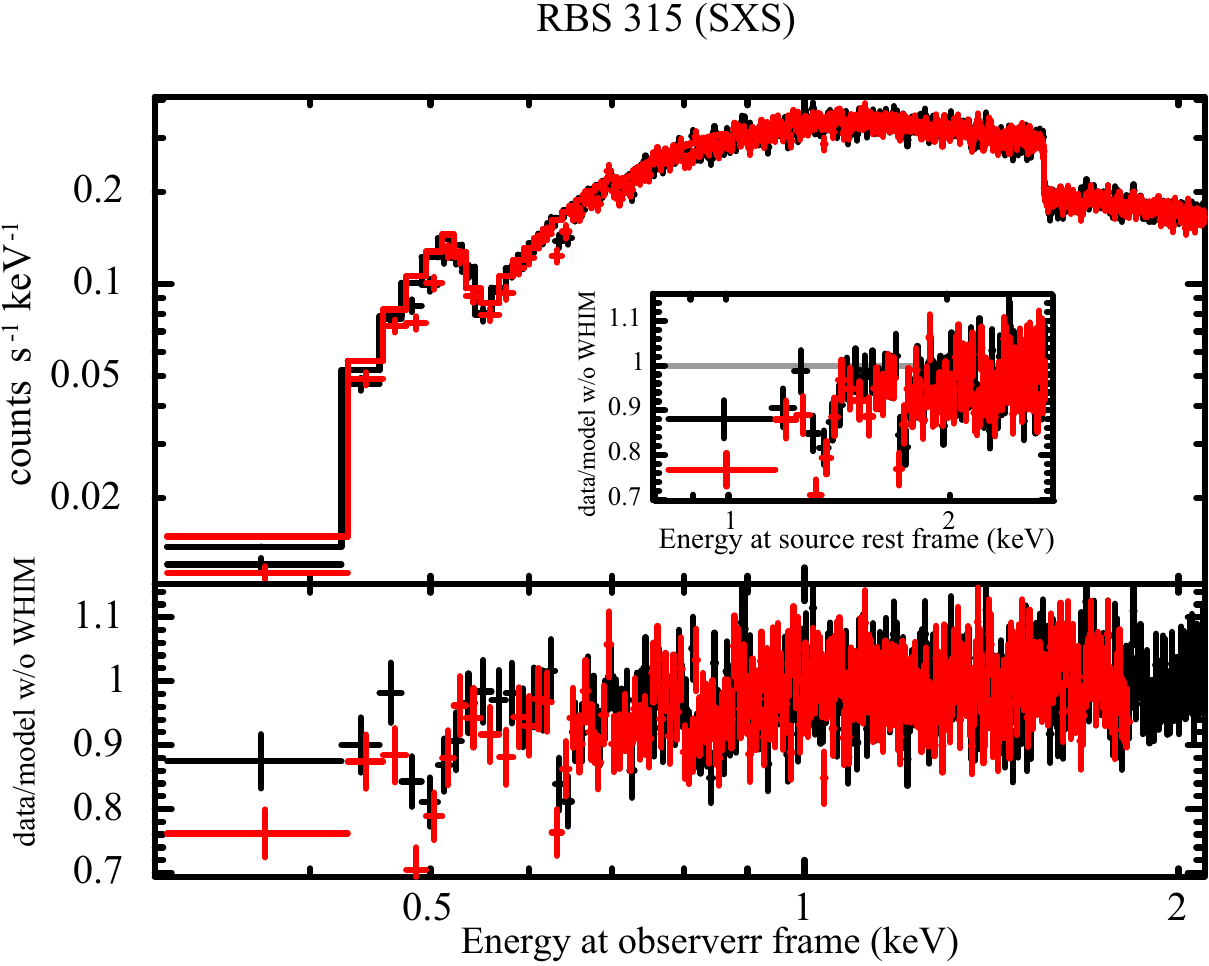}
      \includegraphics[width=7cm]{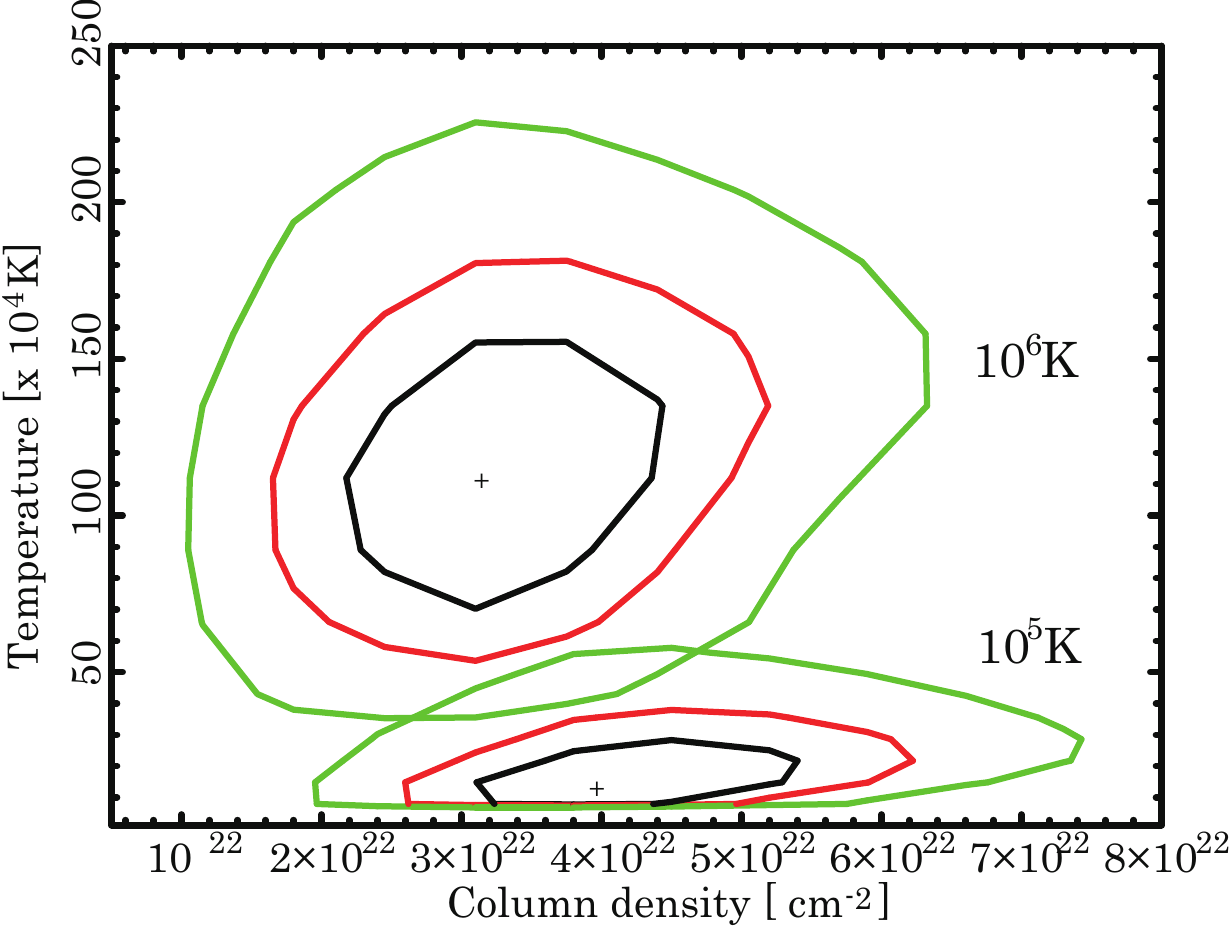}
} 
     \figcaption{(left) The {\it ASTRO-H}/SXS simulated spectrum of the Blazar RBS~315 
(utilizing the model of \cite{Behar11}. The reported intrinsic absorption column 
was replaced with WHIM and Galactic absorption.
Both absorption models are calculated with XSTAR. 
The assumed temperatures of WHIM are $10^5$~K (red) and $10^6$~K (black) . 
Other parameters employed here are the same as those used in \S~3.1 except 
that the WHIM absorber is  redshifted  $(z=2.69)$ to demonstrate the capability determining 
the redshift of absorber. 
The histogram shows the model without the intrinsic absorption but only with 
the Galactic absorption to demonstrate the expected intrinsic absorption features 
in data to model ratios in the bottom panel. 
The inset shows the same features in ratios with energy in the WHIM rest frame. 
The ionization features are clearly shown in the energy band below 1.0 keV.
(right) Confidence contours to demonstrate that SXS will clearly distinguish the assumed 
two different temperatures. 
The WHIM model generated with XSTAR is employed to evaluate each intrinsic absorption 
feature that we assumed in the left simulations.}
\label{fig_RBS315_temp}
\end{figure}

\begin{table}
\tblcaption{Spectral parameters of the Target Blazars \citep{Behar11}}\label{tab:Blazars}
\begin{center}
\begin{tabular}{ccccccl}
\hline \hline 
target name & redshift & $N_{\rm H}^{Gal. {\rm a}}$ &$N_{\rm H}^{z, {\rm b}}$ & flux$^{\rm c}$ 
& reference \\
\hline 
PKS~$2126-158$   &  3.366 & 4.92 & 1.80 & 1.10 & \cite{Fiore03} \\
RBS~315          &  2.690 & 9.26 & 2.90 & 1.08 & \cite{Tavecchio07} \\
PKS~$2149-306$   &  2.345 & 1.61 & 0.08 & 1.00 & \cite{Bianchin09}\\
4C~71.07         &  2.172 & 2.85 & 0.09 & 1.40 & \cite{Malizia00}\\
\hline 
\end{tabular}
\end{center}
\scriptsize{a: Galactic column density in the unit of $10^{20}$H~cm$^{-2}$}\\
\scriptsize{b: Intrinsic column density in the unit of $10^{22}$ H~cm$^{-2}$}\\
\scriptsize{c: Observed 2--10 keV flux in the unit of $10^{-11}$erg~cm$^{-2}$s$^{-1}$}

\end{table}

We show the {\it ASTRO-H} capability in determining the redshift of the possible absorber 
in Figure~\ref{fig_RBS315_redshift}. 
All the simulated spectra shown here assumes the exposure time of 300~ ks.
Thanks to the excellent energy resolution of the SXS crucial for  determining the absorption edge energy 
together with the large effective area for the SXI to determine the optical depth, 
we see that {\it ASTRO-H} will be able to resolve the absorption line structure to determine 
the redshift with X-ray spectrum.

In Figure~\ref{fig_RBS315_temp}, we show that the SXS has a capability to measure 
the temperature of the warm-hot plasmas.
The signature of the  excess absorption is clearly seen in the data to model ratios, 
where we removed the intrinsic WHIM absorption model from the model denoted with histogram 
in Figure~\ref{fig_RBS315_temp}.
In the ratios, we see clear absorption lines due to: 
Si$_{\rm VII - XI}$ at 0.48~keV (1.8~keV at rest frame), 
S$_{\rm VIII - XIII}$ at 0.64~keV (2.4~keV at rest frame), 
and Ar$_{\rm X - XII}$ at 0.81~keV (3.0~keV at rest frame), 
as well as we see in the inset indicated with energy at the source (WHIM) rest frame. 
In addition, the difference of temperature of WHIM is seen in the lowest energy bin in the ratios.
We also show the confidence contours for the two different temperature WHIM.
In the right panel of Figure~\ref{fig_RBS315_temp}, we clearly see that {\it ASTRO-H}/SXS  determines 
the WHIM temperature  with reasonable accuracy.

\begin{figure}[htb]
\centerline{
      \includegraphics[width=0.6\hsize]{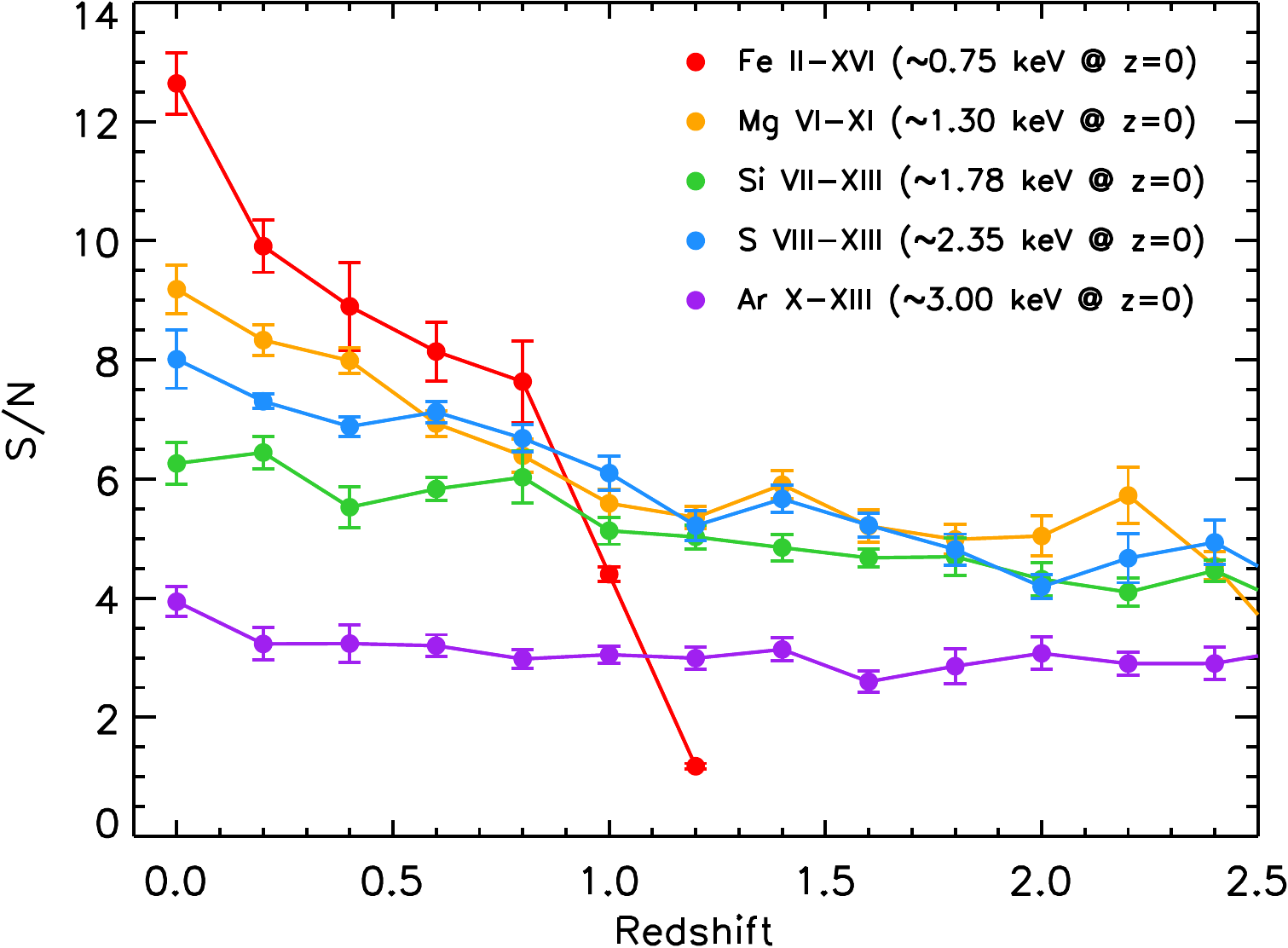}
} 
     \figcaption{The expected significance of each element absorption line of WHIM at $z = 0 - 2.5$
with the 300 ks observation of the {\it ASTRO-H} SXS/SXI.
The Blazar RBS~315 whose X-ray flux and spectral parameters are is described in table 3 is assumed 
to be the WHIM irradiating source. The assumed parameters of WHIM are described in \S~3.1
while the temperature employed in this simulation is $10^5$ K.}
\label{fig_various_z}
\end{figure}

We assumed the WHIM at the same redshift frame with the source and demonstrate that SXS with SXI
are able to determine the column density and the temperature.
Then we simulated the WHIM at various redshift frame from $z = 0$ to $z = 2.5$ to investigate
expected significance of each spectral features of elements. 
Figure~\ref{fig_various_z} shows the signal to noise ratios of absorption lines of Fe, Mg, Si, S, or Ar
of the WHIM at each redshifts. The assumed WHIM temperature is $10^5$ K and other parameters
were the same as described above.
All elements but Ar are to be detected with the significance of $S/N > 4$ at any redshift between 0 and 2.5.
Moreover WHIM in the range of $0 < z < 1$, in which large portion of the WHIM are expected (section 1),
the expected significance fairly exceeds 5 or more.

Since the intrinsic Blazar exhibits featureless spectrum, every bright distant Blazar is  a potential target 
to search for the intervening WHIM and/or other absorption materials.
Many of these sources have strong flaring episodes and are monitored by {\it MAXI} and 
{\it Swift} and thus are candidates for ToO observations. 
In addition to those Blazars listed in table~\ref{tab:Blazars}, a number of Blazars in the redshift range 
from $z= 1 - 3$ have a  high X-ray flux $> 10^{-11}$~erg~cm$^{-2}$~s$^{-1}$ in hard X-ray band 
\citep{Tueller10, Ghisellini11}. 
For example, we could add Blazars Swift~J1656.3$-$3302, [HB89] 0212+735, or [HB89] 0836+710 
for their expected high flux in the SXS band. However, neither Swift~J1656.3$-$3302 sitting behind 
the Galactic center region nor 0836+710 exhibiting rather low intrinsic absorption are suitable for the 
early phase target to explore the WHIM. The rest 0212+735 could be added the target list to be
monitored, although the uncertainty of the reported intrinsic  absorption are fairly large in comparison 
with RBS~315 discussed above. 

\section{Additonal Science --- hard X-ray observations of Blazar and GRB Continua}
As intensive follow up observations with {\it Swift} have revealed, most of GRB afterglows 
experience a series of decay phases, such as a steep decay, a shallow 
decay, a normal decay and a {\it jet decay} phase after the jet break \citep{Nousek06}.
Assuming that  follow-up observation start $\sim 30$ -- 60  hours after the burst, 
as discussed in \S~\ref{sec:GRBnumber}, it should be possible to observe the {\it jet decay}
phase in hard X-rays.
Since the jet break is thought to be caused by deceleration of a shocked shell in the
GRB jet, such measurements will  scan the shock shell from small to large radius as the beaming factor
decreases after the jet break via measuring the  {\it curvature} of the 
continuum using a wide band spectrum as a function of time.  
In other words, observing the jet break is a scanning observation of the shock front, since
  the beaming factor of the emission region decreases dramatically during the jet break decays.
This will provide an excellent chance to resolve the GRB jet internal structure
by utilizing the  wide band {\it ASTRO-H} X-ray data.

While the {\it Suzaku}/HXD has observed four GRB afterglows a few hours after 
the GRB prompt emissions, it did not accurately measure the hard X-ray
spectra because of its limited sensitivity.
The HXI, however, with a two orders of magnitude higher sensitivity, will be
able to observe the hard X-ray spectrum as well as the soft X-ray instruments
(SXI/SXS), as far as the follow up observation by {\it ASTRO-H} is carried out.

The HXI and SGD data will precisely  determine the continuum shape of the afterglow, which increases the accuracy of column density measurement 
discussed above. 

\clearpage
\begin{multicols}{2}
\end{multicols}

\end{document}